\documentclass{iaus}
\usepackage{graphicx,natbib}

  \checkfont{eurm10}
  \iffontfound
    \IfFileExists{upmath.sty}
      {\typeout{^^JFound AMS Euler Roman fonts on the system,
                   using the 'upmath' package.^^J}%
       \usepackage{upmath}}
      {\typeout{^^JFound AMS Euler Roman fonts on the system, but you
                   dont seem to have the}%
       \typeout{'upmath' package installed. iaus.cls can take advantage
                 of these fonts,^^Jif you use 'upmath' package.^^J}%
      }
  \else
  \fi

  \checkfont{msam10}
  \iffontfound
    \IfFileExists{amssymb.sty}
      {\typeout{^^JFound AMS Symbol fonts on the system, using the
                'amssymb' package.^^J}%
       \usepackage{amssymb}%

      }{}
  \fi

  \IfFileExists{amsbsy.sty}
    {\typeout{^^JFound the 'amsbsy' package on the system, using it.^^J}%
     \usepackage{amsbsy}}
    {}

\newsavebox{\astrutbox}
\sbox{\astrutbox}{\rule[-5pt]{0pt}{20pt}}

\title[Outskirts of Galaxy Clusters: intense life in the suburbs]
      {The $K$ Band Luminosity Functions of Galaxies in High Redshift Clusters}
\author[Simon Ellis and Laurence Jones]
{Simon Ellis$^1$
  \thanks{Present address: Anglo-Australian Observatory, PO Box 196, Epping, NSW, Australia},
Laurence Jones$^2$}
\affiliation{$^1$Anglo-Australian Observatory, Australia email: sce@aaoepp.aao.gov.au\\[\affilskip]
$^2$University of Birmingham, UK email: lrj@star.sr.bham.ac.uk}

\pubyear{2004}
\volume{195}
\pagerange{1--8}
\date{?? and in revised form ??}
\jname{Outskirts of Galaxy Clusters: intense life in the suburbs}
\editors{A. Diaferio, ed.}
\begin{document}

\maketitle

\begin{abstract}
$K$ band luminosity functions (LFs) of three, massive, high redshift clusters of galaxies are presented.  The evolution of $K^{*}$, the characteristic magnitude of the LF, is consistent with purely passive evolution, and a redshift of formation $z_{{\rm f}} \approx 1.5$--2.  
\end{abstract}

\firstsection 
\section{Introduction}

Near-infrared photometry of galaxies in high redshift clusters provides an excellent method of studying evolution in the baryonic masses of cluster galaxies.  The $K$ band is rather insensitive to the enhanced blue emission associated with young stellar populations.  This means that observations made in the $K$ band are measuring emission from the old stellar populations, thus the near-infrared (NIR) luminosity is much more closely correlated with the stellar mass of a galaxy than optical luminosities which may be significantly inflated by the presence of only a small, but young, stellar population.  The $K$ band has the added advantages that it has relatively small $k$-corrections, and these are very similar for galaxies of different spectral-type, and the extinction due to dust is also small.   Therefore by measuring the LFs of galaxies in high redshift clusters, and comparing to well established LFs of nearby systems, it is possible to determine how the distribution of stellar mass throughout the cluster population has evolved.  

\section{Sample and Observations}

We have obtained $K$ band photometry of the central regions of three of the most massive ($\sim 10^{14 - 15}$M$_{\odot}$), high redshift ($z=0.83 - 1.03$) clusters of galaxies known.  All three were discovered in the WARPS serendipitous X-ray survey (\citealt{sch97}).  Observations were made with UFTI on the UKIRT telescope on Mauna Kea, Hawaii.  The seeing was typically half an arcsecond in $K$.  The observations typically covered 3 arcminutes square, corresponding to 30\%--40\%
 of the virial radii.  Further details of the observations and analysis can be found in \citet{ell04}.

\section{Modelling Passive Evolution}

In the traditional monolithic collapse picture of galaxy formation (\citealt{egg62}) all the stars in a galaxy are formed in an inital burst, and thereafter the galaxy evolves only in a passive, quiescent manner as its stars make their journey along the main sequence.  This results in a gradual dimming of the stars, and consequently the galaxy.  Any further episodes in the evolution of galaxies, such as secondary bursts of (or gradual) star-formation activity, mergers, and gravitional effects such as harrassment, stripping, starvation, etc.\ must take place against this background of the  passive evolution of the pre-existing stars.

Thus in order to establish whether any of these `extra-passive' processes are important in the evolution of cluster galaxies (or more strictly, if they are important in the evolution of the $K$ band luminosities of cluster galaxies) we adopt the approach of modelling a passively evolving system, and searching for irregularities between the predictions of the model and observations.  Unless the $K$ band is insensitive to these extra-passive events they should show up as differences.  Thus it may be expected that the following analysis is more senstive to issues regarding the mass-assembly of the galaxies, than it is to the star-formation history of galaxies.

We have modelled passive evolution using the synthetic stellar population libraries of \citet{bru03}.  Stellar populations with a Salpeter initial mass function and solar metallicity were assumed.  The evolution was modelled for instantaneous bursts at $z_{f}=$1.5, 2 and 5, along with predictions of no evolution.

\section{Results}

\begin{figure}
\begin{minipage}[c]{0.5\textwidth}
    \centering \includegraphics[scale=0.26,angle=270]
    {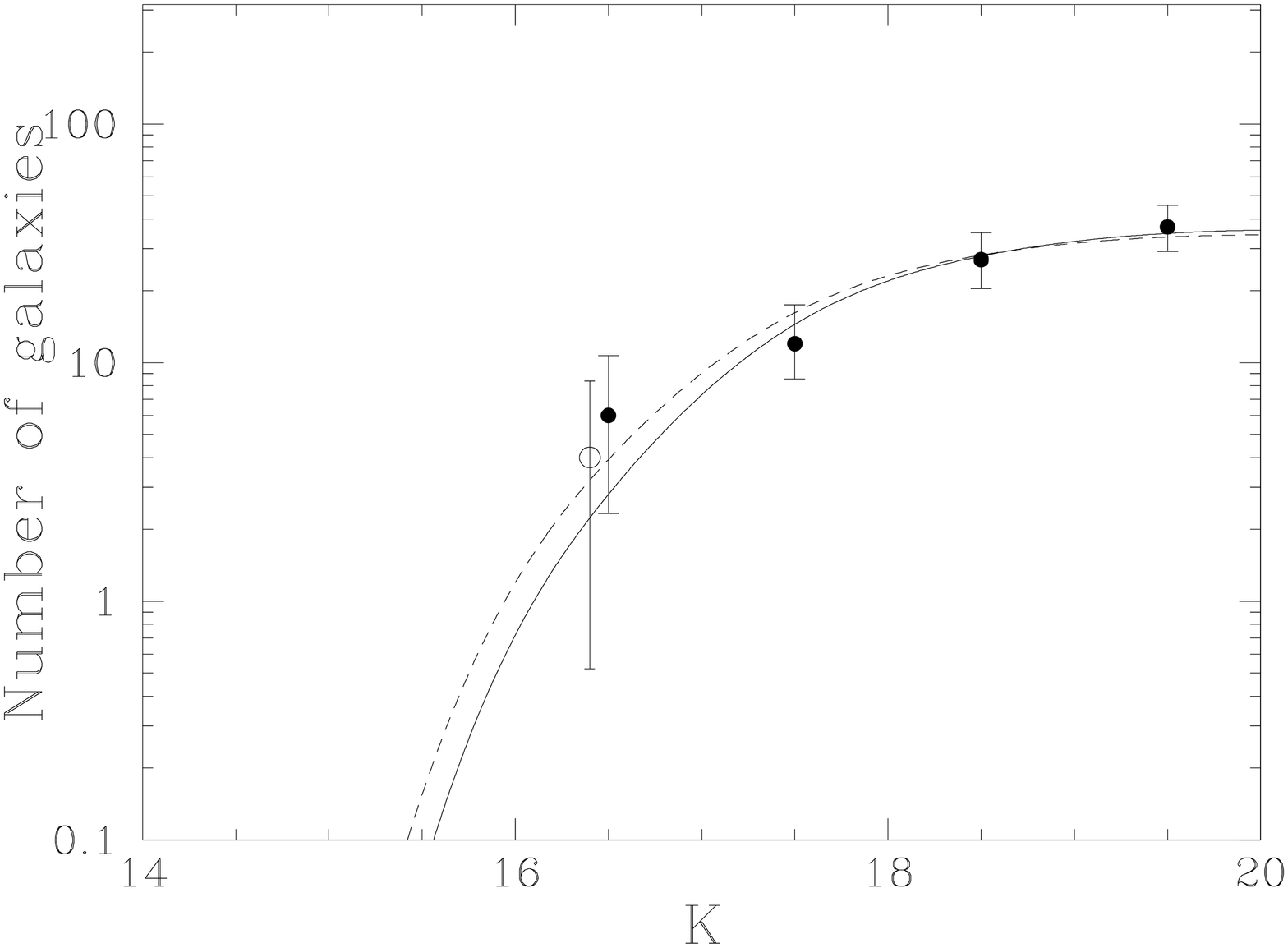}
  \end{minipage}%
  \begin{minipage}[c]{0.5\textwidth}
    \centering \includegraphics[scale=0.26,angle=270]
    {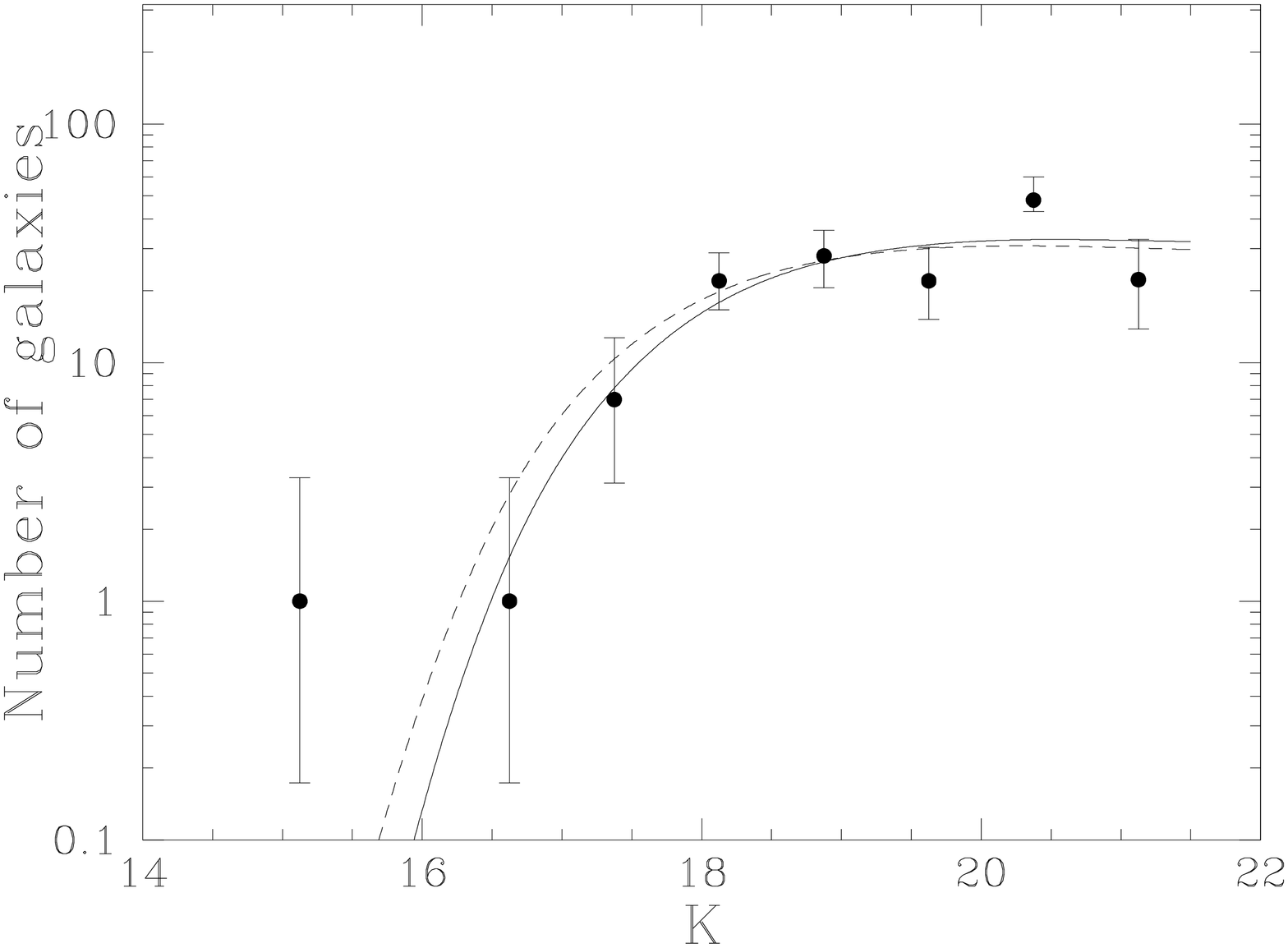}
  \end{minipage}
  \begin{minipage}[c]{0.5\textwidth}
    \centering \includegraphics[scale=0.26,angle=270]
    {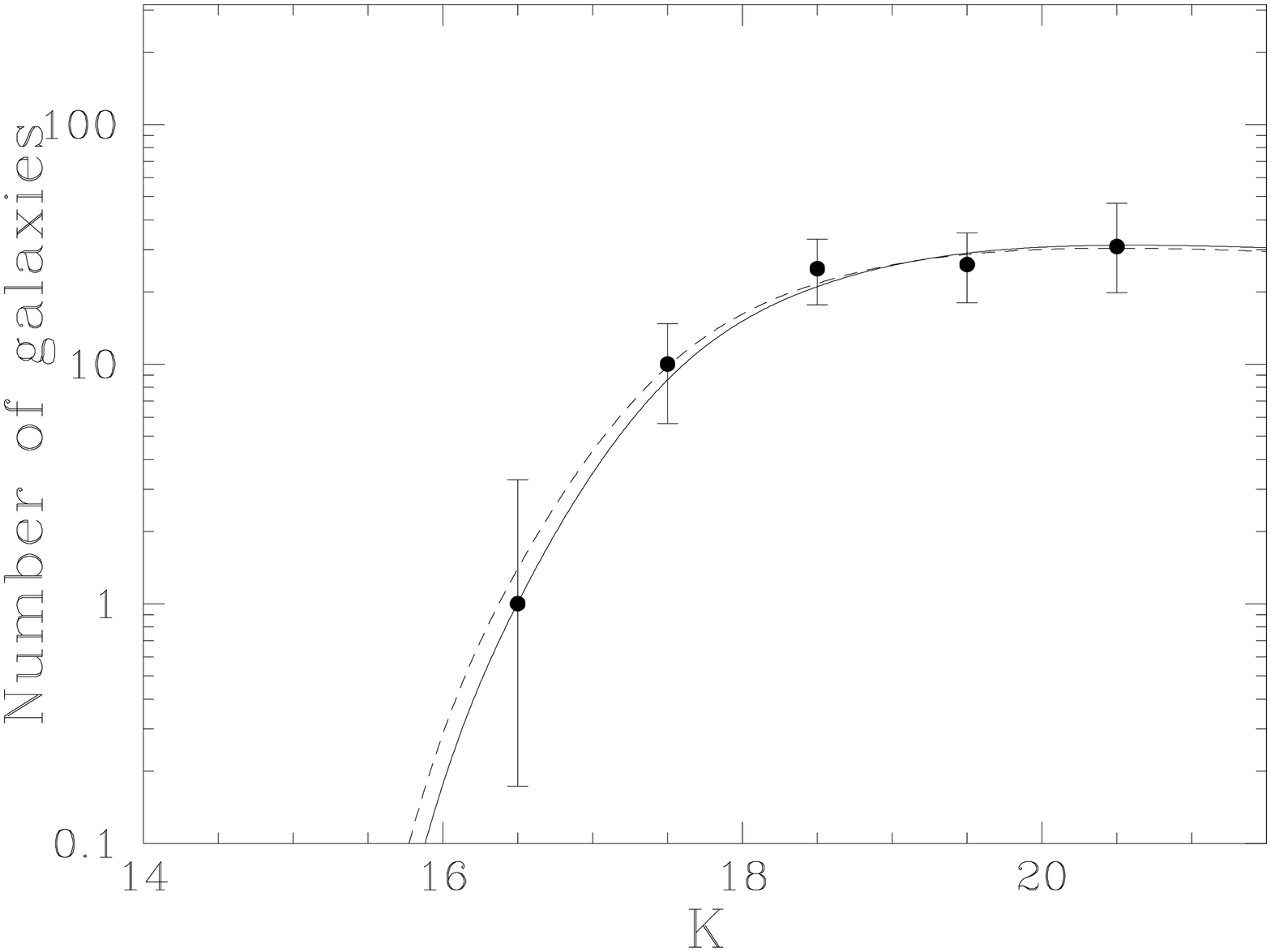}
  \end{minipage}%
  \begin{minipage}[c]{0.5\textwidth}
    \centering \includegraphics[scale=0.26,angle=270]
    {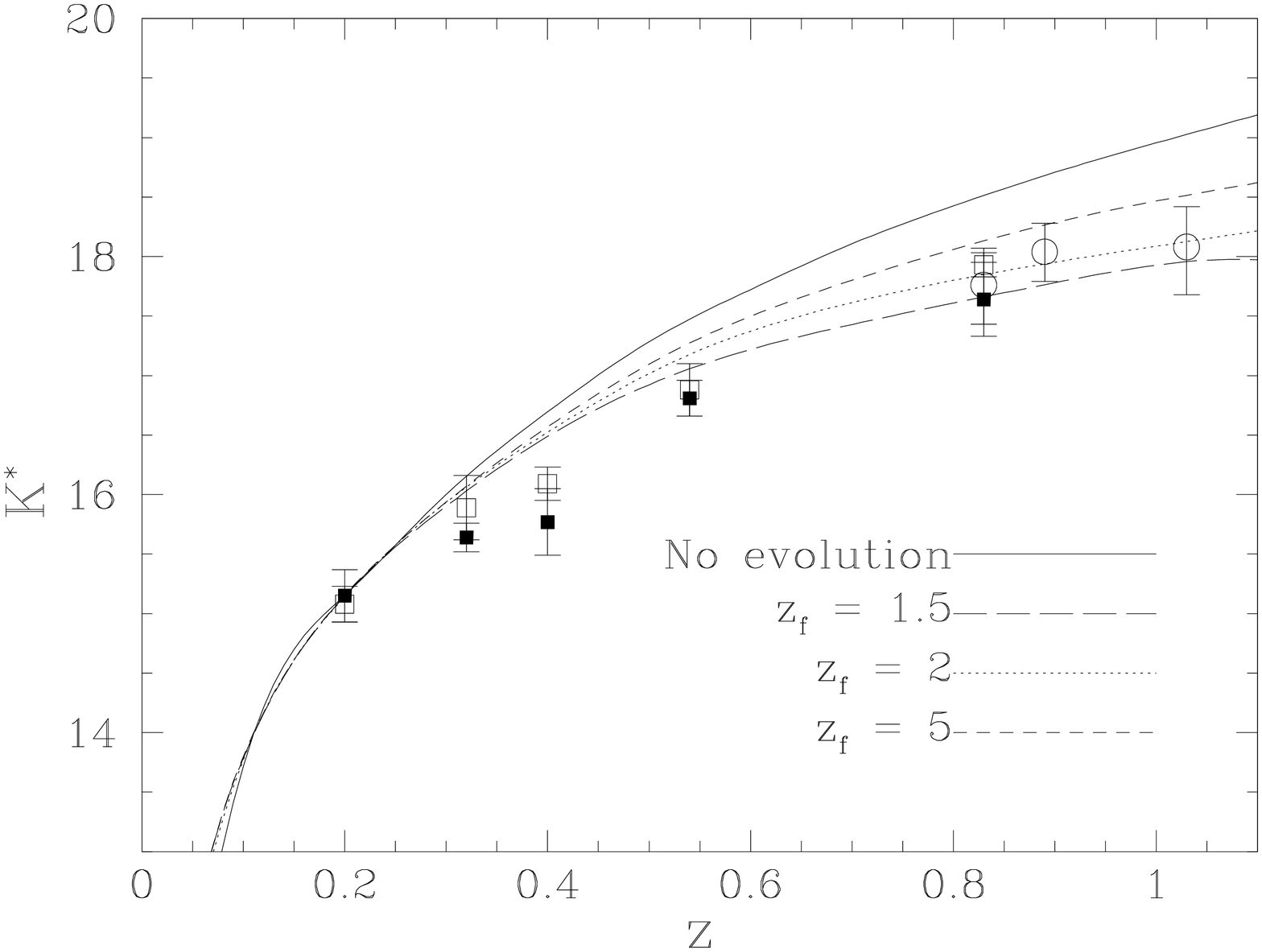}
  \end{minipage}
\caption{The evolution of the $K$ band LF.  The first three panels show the LFs of the clusters of galaxies individually: top-left, ClJ0152; top-right, ClJ1226; bottom-left, ClJ1415.  The lower right-hand quadrant shows the evolution of $K^{*}$.
The circles are data from this paper.  The squares are from \protect\citet{dep99}, open symbols being low $L_{{\rm X}}$ systems and closed symbols being high $L_{{\rm X}}$ systems.}
\label{fig:kstar}
\end{figure}

LFs were constructed individually for each cluster after applying a statistical subtraction of foreground/ background galaxies based on number counts of galaxies from nearby offset fields observed at the same time as the clusters, and on spectroscopic information for the brighter cluster members, where available.  The LFs for ClJ0152 (top left), ClJ1226 (top right) and ClJ1415 (bottom left) are shown in figure~\ref{fig:kstar}.

\citet{sch76} functions were fit to the LFs of each cluster individually, with the faint end slope fixed at $\alpha=-0.9$ (the local value for the Coma cluster, \citealt{dep98}, and the field, \citealt{gar97}), since the data were not deep enough to constrain both $\alpha$ and $K^{*}$ precisely.  The evolution of $K^{*}$ with redshift is shown in the lower right panel of figure~\ref{fig:kstar}, along with lower redshift points from \citet{dep99}.  The models (described in the previous section) are normalised at low redshift to $K^{*}$ of the Coma cluster (\citealt{dep98}).  The evolution is seen to be consistent with passive evolution models, with $z_{{\rm f}} \approx 1.5$ if the models are normalised to Coma, or a higher redshift of formation if the models are normalised to the low redshift points from \citet{dep99}.

Purely passive evolution of early-type galaxies is consistent with several other
 studies including the evolution of the $K$ band luminosity 
function (\protect\citealt{dep99}, \citealt{tof03}), studies of 
the scatter of the colour-magnitude relation 
(see e.g.\ \protect\citealt{ell97}, \protect\citealt{sta98}) and
evolution of the fundamental plane in terms of mass-light ratios
(\protect\citealt{van98}).

Whilst the observed evolution is fully consistent with passive evolution, it may be reconciled with the framework of hierarchical formation of structure, so long as any merging between luminous galaxies in massive clusters takes place before $z\approx1$.  As the systems under consideration here are unusually massive, they will be associated with regions of large over-density, and thus it is perhaps expected that objects will assemble collapse at early epochs in such an environment.  Furthermore, the large velocity dispersions associated with massive clusters makes the likelihood of mergers small (see eg.\ \citealt{dia01}).  For a more thorough discussion of this work see \citet{ell04}. 


\begin{acknowledgments}
The authors would like to thank warmly both Ben Maughan and Harald
Ebeling. UKIRT staff have been
very efficient and helpful; service time
observations were performed by some of them.  
The authors would also like to thank 
Stefano Andreon and Antonaldo Diaferio for their help.  
\end{acknowledgments}
\bibliographystyle{mn2e}
\bibliography{clusters}





\end{document}